\documentclass[twocolumn]{aastex62}

\newcommand{\RSun}{\ensuremath{R_\odot}}

\graphicspath{{./}{figures/}}

\shorttitle{}
\shortauthors{Sarkar et al.}

\begin{document}

\title{\textbf {\large Integration of a CME Flux Rope Model with Magnetic Field Draping in Heliospheric Simulations}}

\correspondingauthor{Ranadeep Sarkar}
\email{ranadeep.sarkar@helsinki.fi}

\author{Ranadeep Sarkar}

\author{Jens Pomoell}

\author{Emilia Kilpua}

\author{Eleanna Asvestari}

\affiliation{Department of Physics, University of Helsinki, Helsinki, Finland} 

\begin{abstract}

Accurate prediction of the magnetic structure of interplanetary coronal mass ejections (ICMEs) remains a key challenge in space weather forecasting. Despite several efforts using global MHD modeling, it remains difficult to deduce a flux-rope solution that can accurately and reliably reproduce the magnetic structure of a CME. We present FROST (Flux Rope embedded in the Solar Magnetic Field), a new flux-rope model implementation within the European Heliospheric Forecast Information Asset (EUHFORIA), featuring an initially force-free toroidal flux rope embedded in the low-coronal magnetic field. FROST employs a novel embedding technique that self-consistently generates a draping magnetic field around the flux-rope, ensuring continuity of the normal magnetic field component at the CME boundary. The dynamics of the flux-rope in the low and middle corona is solved by a non-uniform advection constrained by the observed kinematics of the event. This results in a global, non-toroidal, stretched loop-like magnetic structure. At heliospheric distances, the evolution is modeled as an MHD process using EUHFORIA. We validate the FROST model implementation by simulating the CME event of 2013 April 11 and comparing the results with in situ magnetic field measurements at 1 au. The modeled magnetic field profiles of the flux-rope exhibit strong agreement with the observations, with correlation coefficients ranging from 0.64–0.93 for $B_x$, 0.92–0.95 for $B_y$, and 0.82–0.87 for $B_z$ component of the magnetic field. These results demonstrate the capability of the proposed flux-rope model to significantly improve ICME magnetic field prediction, highlighting its strong potential for advancing space-weather forecasting capabilities.

\end{abstract}


\section{Introduction} \label{sec:intro}
Coronal mass ejections \citep[CMEs; e.g.,][]{Webb2012} are  massive clouds of magnetized plasma that frequently erupt from the Sun and are one of the major sources of space weather disturbances \citep[e.g.,][]{Temmer2021,Kilpua2017}. When a CME travels significantly faster than the ambient solar wind, a shock wave forms ahead of the CME ejecta. The turbulent region of hot and compressed plasma between the shock front and the CME ejecta is known as the sheath region. Depending on the solar wind density, speed, and more critically the strength and duration of the southward component ($B_z$) of the interplanetary magnetic field, a CME or its sheath region can produce a severe geomagnetic storm upon impacting Earth’s magnetosphere \citep{Wilson_1987,tsurutani_1988,Gonzalez_1999,Huttunen_2005, Gopalswamy_2008}. Therefore, it is crucial to accurately model the Sun-to-Earth evolution of magnetic field vectors within interplanetary CMEs \citep[ICMEs; e.g.,][]{Kilpua2017LRSP} and its sheath region based on near-Sun observational proxies, to get advance alert about their potential geo-effectiveness.

\begin{figure*}[t!]
\begin{center}
\includegraphics[width=\textwidth,clip=]{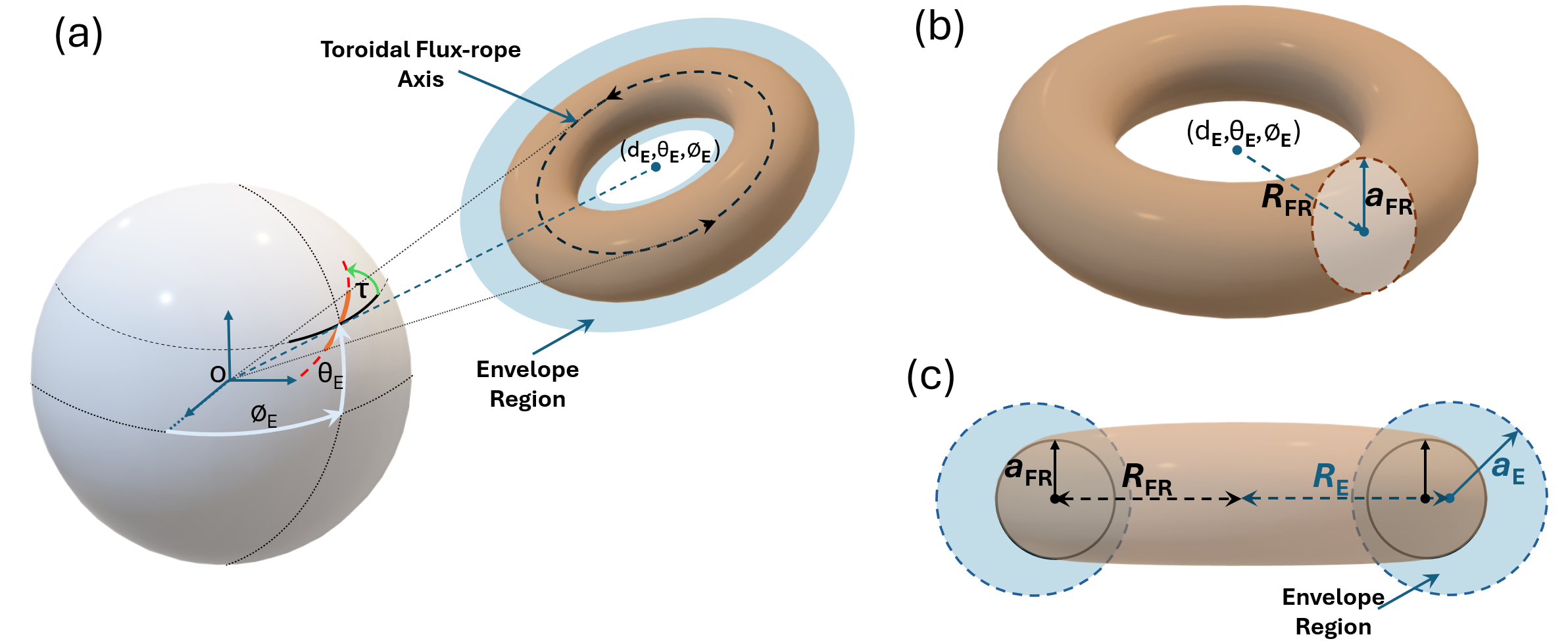}
\caption{Schematic diagram of the toroidal embedding region containing the flux rope (brown solid torus) and the surrounding envelope region (blue). Panel (a) shows the 3D orientation of the torus with tilt angle ($\tau_E$), with its center located at $(d_E, \theta_E, \phi_E)$ in the HEEQ coordinate system, where O denotes the Sun-center. Panels (b) and (c) show an overview of the cross-sectional geometry of the torus.}\label{torus_cartoon}
\end{center}
\end{figure*}

Modeling the interplanetary evolution of CMEs is particularly challenging due to the lack of observationally constrained flux rope solutions that can reliably capture their internal magnetic configuration and evolution \citep{TEMMER2023}. Many global heliospheric MHD modeling approaches commonly use an axi-symmetric spheromak or modified spheromak configurations to characterize the magnetic structure of a CME and simulate its evolution from the Sun to Earth \citep[e.g.][]{1998Vandas,2002Vandas,Manchester_2004,Shiota2016,2019scolini,Scolini_2020, Asvestari1, Singh_2022}.

However, fully inserted spheromaks are inconsistent with the canonical scenario in which the two legs of a CME remain connected to the Sun \citep{Gosling_1987,farrugia_1995,Owens_2016}. This scenario is supported by observations of bi-directional electrons; and even though spheromaks can form naturally in magnetized environments they do not align with this observational aspect. Consequently, their lack of leg-like components renders them not suitable for representing ICMEs encountered by a spacecraft away from their apex. Additionally, even for ICMEs that are cut through or close to their apex, fully inserted spheromaks result in an unrealistic double flux rope signature when compared with in-situ spacecraft observations \citep[e.g.,][]{Sarkar_2024}. 

More importantly, a spheromak injected into the inner heliospheric domain undergoes evolutionary processes that cause it to rotate and/or deflect \citep{Asvestari2}. While CMEs may undergo rotation or deflection due to its interaction with the ambient solar wind, the injection of a spheromak model can introduce additional rotation and deflection. This may appear unrealistic compared to the behavior expected from the standard canonical configuration of a CME as mentioned above. Therefore, the aforementioned discrepancy can severely affect the predictive accuracy of space weather forecasting models when applied to CME events that actually feature such an anchored configuration \citep{Sarkar_2024}.

Implementation of flux rope (FR) models \citep[][]{Michigan_group,PSI,Singh_2022,MAHARANA2022,Gamera,Maharana_2024}, where the FR legs remain attached to the inner boundary of the simulation domain, overcomes the limitation of fully inserted spheromaks as pointed out earlier. Notably, current state-of-the-art heliospheric models either superimpose a flux rope solution onto the background heliospheric environment or insert it via the inner boundary, leading to unphysical discontinuities in the magnetic field at the insertion location on the boundary or unrealistic pressure non-equilibrium. Additionally, the aforementioned FR implementation techniques may hinder the accurate and self-consistent formation of sheath regions close to the FR boundary. Even when modeling CMEs as unmagnetized transients, the precise prescription of the evolution of the plasma at the heliospheric surface through which the transient is introduced into the simulation affects significantly the prediction of geoeffectiveness \citep{scolini_2018}.  

In this work, we present FROST (Flux Rope embedded in the Solar Magnetic Field), a new FR model implementation for global MHD simulations, aiming to mitigate the limitations of existing CME models for space-weather science and forecasting. In section \ref{methods}, we first describe the method of embedding the FR into the coronal domain (below 0.1 au) and its insertion into the heliospheric domain (above 0.1 au) of the EUropean Heliospheric FORecasting Information Asset (EUHFORIA) \citep{2018Pomoell}. The heliospheric evolution of the CME is simulated with EUHFORIA and we validate this new approach in section \ref{validation}, using a well-observed CME event that erupted on 2013 April 11. In Section \ref{validation}, we present a detailed description of the method used to realistically constrain the flux rope using remote sensing solar observations, along with a comparison between the magnetic field and other plasma property time series simulated with EUHFORIA and those observed at Earth's orbit. Finally, we summarize our results and conclude in section \ref{conclusion}. 

\begin{figure*}[t!]
\begin{center}
\includegraphics[width=\textwidth,clip=]{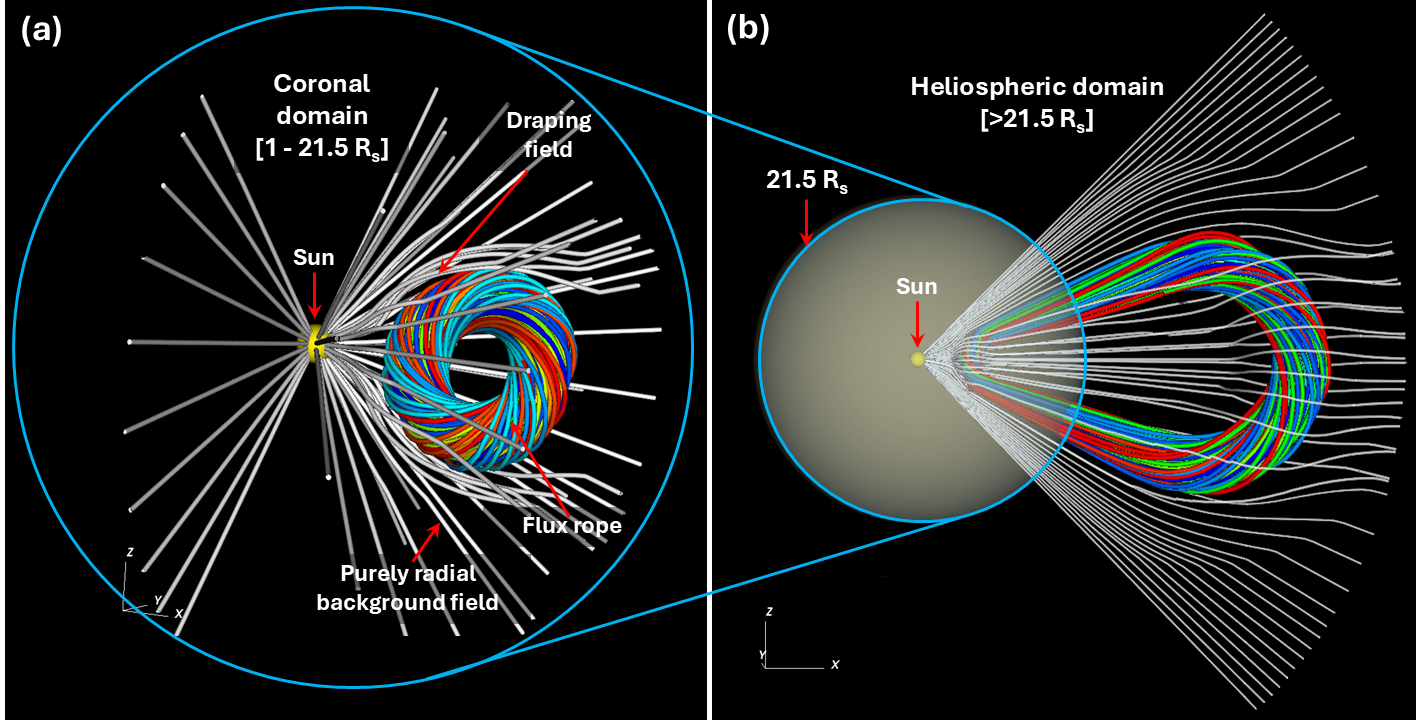}
\caption{(a) Toroidal flux rope embedded in the coronal magnetic field. The background coronal magnetic field, including the draping field around flux rope boundary, is shown by gray solid curves, while the color-coded, twisted field lines depict the toroidal flux rope. (b) The magnetic structure of the flux rope (multi-color field-lines) at an advanced time after applying the non-uniform advection profile which dominates only at the frontal half of the torus. The background magnetic field close to the $y=0$ plane within $45^{\circ}$ latitude with respect to the Sun-Earth line ($x$-axis) is shown by the gray solid curves. }\label{advected_torus}
\end{center}
\end{figure*}

\section{Methods}\label{methods}
The basic principle of introducing a transient CME structure into the ambient heliosphere characterized by the solar wind follows the same approach as in previous dynamic heliospheric modeling using magnetized or unmagnetized models \citep{2018Pomoell,2019scolini}. In this approach, the CME is gradually injected into the heliospheric domain using a time-dependent prescription at the radial inner surface of the model at $r = R_0 = 0.1$ au. However, in contrast to previous work, the FROST modeling approach injects the magnetic field of the flux rope, including a surrounding volume. The primary aim of this novel approach is to include the draping of the ambient magnetic field around the flux rope, ensuring a consistent transition from the surrounding solar wind magnetic field to that of the ejecta and the disturbed surrounding region. Furthermore, in this process, constraints arising from the mathematical structure of the magnetic field are respected in that the normal component of the magnetic field is continuous across all surfaces, as required for the magnetic field to be divergence-free. In addition, the insertion is constructed so that the legs of the FR become stretched and less twisted with time, as expected for the canonical large-scale structure of FRs (as discussed in Section 1).

To achieve the above, the construction of the magnetic field of the CME to be injected needs to be generalized to take into account the surrounding magnetic field of the ambient solar wind. This is accomplished in two steps, where in the first step the magnetic field of the entire ejecta, including the flux rope magnetic field and its surroundings, is first constructed in the coronal domain (i.e., below the domain of the heliosphere,  $r \in [\RSun, R_0]$). In a second step, the constructed magnetic field is transported in time to $r=R_0$ providing the plasma and magnetic field at the inner radial boundary surface of the heliospheric model. In the following subsections, the two distinct steps of the procedure are explained in further detail. 

\subsection{Embedding of a magnetized flux rope in an external field}
\label{subsec:embedding}

Constructing the FROST model that provides the magnetic field and plasma environment of the ejected structure is carried out by defining a closed volume that fully contains it, denoted henceforth as the embedding region. Outside the embedding region, the magnetic field and plasma conditions represent the ambient undisturbed solar wind, henceforth denoted as the external region. The embedding region is chosen to be a solid torus, with the size of the volume defined by the minor, $a_E$, and the major, $R_E$, radii of the torus, and the location of its center specified using the spherical representation in the Heliocentric Earth Equatorial (HEEQ) coordinate system and denoted by the coordinates $(d_E, \theta_E, \phi_E)$. Contextually, within the HEEQ coordinate system, the heliographic longitude ($\phi$) is measured from the Sun-Earth line, and the heliographic latitude ($\theta$) is measured relative to the solar equator, being positive toward the solar north.

The toroidal embedding region itself is composed of two regions, an inner toroidal volume (minor radius $a_\mathrm{FR}$ and major radius $R_\mathrm{FR}$) that defines the volume occupied by the flux rope, and an envelope region that fills the rest of the embedding region, thus defining a volume between the external region (the undisturbed solar wind) and the toroidal flux rope (see Figure \ref{torus_cartoon}b \& c). The orientation of the torus is restricted so that its axis of symmetry is perpendicular to the position vector from the center of the Sun to the torus. As such, the orientation of the torus is uniquely defined by a single angle, the tilt angle $\tau_E$. The orange curve in Figure \ref{torus_cartoon}a shows the projected magnetic axis (a dashed black circle indicated by the blue arrow in Figure \ref{torus_cartoon}a) of the toroidal flux-rope on a Sun-centered spherical surface. The angle $\tau_E$ is defined as the counterclockwise angular separation between this orange curve and the constant $\theta_E$ arch as marked by the black solid curve.


\subsubsection{Magnetic field in the envelope region}

The function of the envelope region is to act as a region between the undisturbed ambient external region and the flux rope. 
In this work, we assume that the magnetic field in the envelope is given by a potential field, $\mathbf{B}_\mathrm{env} = \nabla u$. At the boundaries of the domain we employ Neumann boundary conditions, i.e., at the toroidal surface between the envelope and the external field, the normal component of the magnetic field is required to be continuous and equal to that of the external field, while at the interface between the envelope and the flux rope, the normal component is set equal to zero, $B_n = 0$. These boundary conditions ensure that the magnetic field in the envelope seamlessly connects to the external field, while the flux rope is magnetically isolated from the surroundings with no flux passing through the toroidal surface between the envelope and the flux rope.

By making use of the inherent symmetry of the toroidal volume, the resulting Laplace equation for the magnetic scalar potential $\nabla^2 u = 0$ can efficiently be solved employing toroidal coordinates. We utilize a spectral decomposition in the periodic toroidal direction, while each poloidal plane is solved using a finite-difference discretization. This combination allows an exactly (i.e., to floating-point precision) divergence-free magnetic field in the chosen staggered discretization to be obtained.

\subsubsection{The flux rope magnetic field}

Due to the inclusion of the envelope region, the toroidal volume at the core of the embedding region is magnetically isolated from the surroundings. Thus, the magnetic field in this region can be populated using any toroidal flux rope model of choice as long as the normal component $B_n = 0$ is respected by the chosen model at the outer edge of the flux rope domain.
%
%

The toroidal flux rope solution that we use in this work to model the magnetic field at the core of the CME is based on the force-free Soloviev solution, which is an exact analytical solution of the Grad–Shafranov equation \citep{1968Soloviev}. The flux rope model is constructed in a local cylindrical $(\rho, \phi, z)$ system and the fields are assumed to be invariant about the azimuthal angle $\phi$, i.e., $\partial_\phi=0$. The magnetic field is given by $\mathbf{B}_\mathrm{FR} = \frac{1}{\rho} [\nabla \psi(\rho, z)] \times \mathbf{e}_\phi + B_\phi(\rho, z) \mathbf{e}_\phi$ where $\psi$ is the magnetic flux function. The force balance $\mathbf{J} \times \mathbf{B} = \nabla p$, with $p$ the thermal pressure, then reduces to the Grad-Shafranov equation
\begin{eqnarray}
    \nabla_*^2 \psi = -\mu_0 \rho^2 \frac{dp}{d\psi} - \mu_0^2 F \frac{dF}{d\psi} 
\end{eqnarray}
where $\mu_0 F(\psi) = \rho B_\phi$ and $\nabla_*^2$ is the Grad-Shafranov operator. The Soloviev solution is obtained for the case when $p$ and $F$ are required to satisfy $\mu_0 \frac{dp}{d\psi} = -c_1$ and $\mu_0^2 F\frac{dF}{d\psi} = -c_2$, where $c_1$ and $c_2$ are constants. The resulting Grad-Shafranov equation can then be solved analytically. In practice, we employ the procedure detailed in \citet{Cerfon2010} to construct the Soloviev solution that conforms to the circular cross-section assumed for the flux rope. 

The heliospheric propagation of a Soloviev flux rope has recently been studied using EUHFORIA \citep{2024Linan}. However, \citet{2024Linan}  considered a full insertion of the Soloviev flux rope into the heliospheric domain, which does not account for scenarios where the two legs of the flux rope remain anchored to the Sun. In addition, an envelope field was not considered in \citet{2024Linan}.


\subsubsection{The external magnetic field}

Outside the embedding volume, the magnetic field in the external region is assumed to represent the ambient solar wind magnetic field. In a typical heliospheric simulation, the ambient solar wind magnetic field in the domain below 0.1 au is obtained using the potential field source
surface (PFSS) and Schatten current sheet (SCS) models \citep{mcgregor}, which can be readily used to define the external field in FROST. However, at heliocentric distances where the embedding of the flux rope is performed, the magnetic field is predominantly purely radial. Therefore, as a simplification, we use as the external magnetic field $\mathbf{B}_\mathrm{ex} = B_0 (\theta, \phi) (R_0/r)^2 \mathbf{e}_r$. 

It is important to note that the assumption of a potential magnetic field in the envelope region is consistent only when the normal component (with respect to the toroidal surface of the embedding) of the current density carried by the external magnetic field at the interface between the envelope and external field is zero. In situations where the ejecta is to be embedded close or within the heliospheric current sheet, this assumptions is not valid. Nevertheless, in this situation, the embedding can be carried out by setting the external field to be unipolar, $\mathbf{B}_\mathrm{ex} = |B_0 (\theta, \phi) | (R_0/r)^2 \mathbf{e}_r$, and after the embedding has been performed, the polarity of the magnetic field can be switched where necessary.

We note that a conceptually similar approach of embedding a flux rope in a surrounding potential field was considered by \citet{Vandas2003}. However, in our approach, the effect of the embedding is local and confined to a given (toroidal) volume. Furthermore, the external field can in our approach be arbitrarily complex, as long as the interface conditions are met. 

The result of embedding the flux rope in an external radial field in the coronal domain is depicted in panel (a) in Figure \ref{advected_torus}. The colored curves show field lines of the magnetically isolated flux rope, while gray curves depict the magnetic field outside the flux rope. Field lines that reach close to the flux rope can be seen to smoothly drape around the flux rope in the envelope region. At the transition between the external and the envelope field, the perpendicular components of the magnetic field can be discontinuous. This is visible in Figure \ref{advected_torus}a in particular at the leading edge where the field lines are seen to rapidly deflect when entering the envelope. 

\subsubsection{Plasma conditions in the embedding region}

In addition to the magnetic field, also the plasma parameters need to be specified. In this work, we assume uniform plasma density and temperature inside the embedding region (the flux rope and surrounding envelope) and assign the user-defined CME mass density and temperature to the volume. Depending on the embedding of the flux rope at a certain coronal height, the average plasma density of the CME can be determined from observational or empirical methods \citep{2021Mannuela}.


\begin{figure*}[t!]
\begin{center}
\includegraphics[width=\textwidth,clip=]{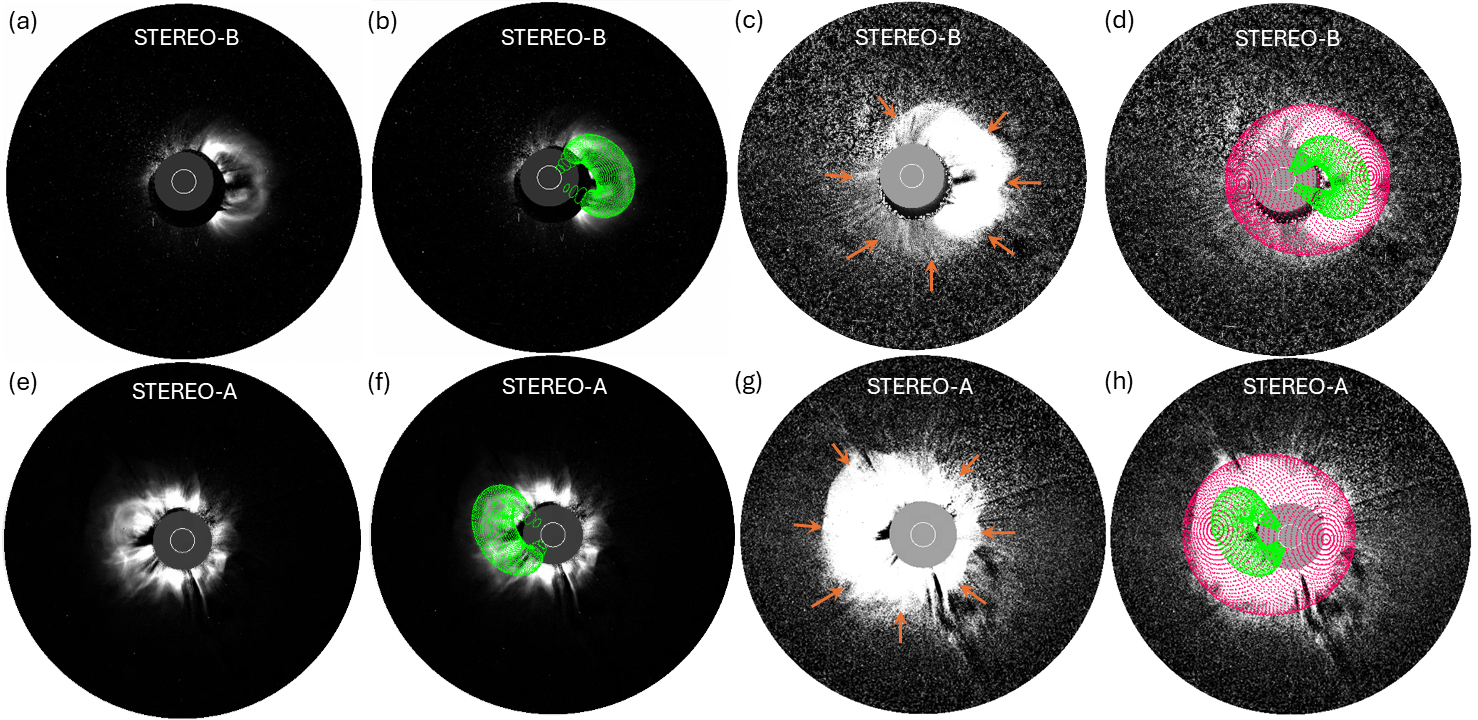}
\caption{CME morphology as observed in background subtracted images of COR2-B (a) and COR2-A (e), respectively, at 08:24 UT on
2013 April 11. Panels (b) and (f) illustrate the over-plot of the fitted wire-frame (green dots) using the GCS model. Panels (c) and (g) are the same as panels (a) and (e), respectively, but the images are over-saturated to highlight the shock boundary, as indicated by the orange arrows. Panels (d) and (h) show the over-plot of the  ellipsoidal wire-frame of the shock (red dots), along with the GCS structure as fitted in panels (b) and (f)}\label{shock_fitting}
\end{center}
\end{figure*}
\subsection{Propagating the embedding region}

The construction of the embedding, detailed in Section \ref{subsec:embedding}, is performed in the external undisturbed ambient solar wind conditions in the coronal domain. Therefore, in order to introduce the ejecta into the heliospheric domain, the solution needs to be propagated to the inner radial boundary surface of the heliospheric simulation. In order to mimic a propagation in which the legs of the flux rope remain anchored in the lower corona, thereby stretching the legs of the flux rope towards a radial field, we employ a propagation scheme that incorporates this effect.


To this end, we consider the case of solving the evolution of the magnetic field under the influence of a radial flow 
\begin{eqnarray}
\mathbf{v}(\mathbf{x}, t) = v_r(r, \theta, \phi, t) \mathbf{e}_r .
\end{eqnarray}
Writing Faraday's law
%
%
in terms of the vector potential $\mathbf{B} = \nabla \times \mathbf{A}$ gives
\begin{eqnarray}
\partial_t \mathbf{A} = - \mathbf{E}. 
\end{eqnarray}
%
A particularly suitable gauge fixing for the vector potential is to employ the so-called multipolar or Poincar\'e gauge, which in the solar coronal context is often referred to as the DeVore gauge after \citet{DeVore2000}. In a spherical domain, this gauge condition is given by $A_r = 0$ \citep[e.g.,][]{Moraitis2018, Amari2013, YeatesHornig2016} and is used here.

With the ideal MHD electric field $\mathbf{E} = - \mathbf{v} \times \mathbf{B}$ and using the assumption of radial flow, the induction equation becomes 
\begin{eqnarray}
\frac{\partial A_\theta}{\partial t} 
 &=& - v_r (\nabla \times \mathbf{A})_\phi 
  = - \frac{v_r}{r}  \frac{\partial}{\partial r} (r A_\theta) \\
\frac{\partial A_\phi}{\partial t} 
 &=& v_r (\nabla \times \mathbf{A})_\theta
  = - \frac{v_r}{r} \frac{\partial}{\partial r} (r A_\phi).
\end{eqnarray}
Therefore, solving the induction equation in this case is equivalent to solving two advection equations
\begin{eqnarray}
\frac{\partial q}{\partial t} &=& - v_r  \frac{\partial q}{\partial r} 
\end{eqnarray}
for $q = r A_\theta$ and $q = r A_\phi$, and can be carried out using the method of characteristics. We note that as a result of employing a staggered discretization for the magnetic field when constructing the embedding, the vector potential components $A_\theta$ and $A_\phi$ that are consistent with the input magnetic field up to floating-point accuracy can readily be computed \citep{Valori2016, Moraitis2018}. We also note that due to the radial flow the radial electric field $E_r = 0$ so that $\partial_t A_r = 0$. Therefore, $A_r=0$ throughout the computation.

The radial flow field is chosen to be given by
\begin{eqnarray} \label{eq:vflow}
v_r = f(r) \frac{r}{t + r_0/v_0}    
\end{eqnarray}
where $f(r)$ is a tapering function that is constant for $r \gtrsim d_E$ and smoothly transitions the flow speed towards zero in the region $r < d_E$ implemented using a hyperbolic tangent. For the case that $f(r)$ is a constant, which is approximately valid in the region $r \gtrsim d_E$, the flow is equivalent to the inertial self-similar velocity flow field \citep{Low1982, GibsonLow1998}. Therefore, the flow field given by Eq. \ref{eq:vflow} provides self-similar expansion of the leading half of the embedding, while keeping the latter half approximately fixed. As a result, the magnetic field of the legs are stretched, mimicking the anchoring of the foot points of the flux rope in the lower corona.

In Figure \ref{advected_torus}b, the result of advecting the magnetic field into the heliosphere using as starting point the solution provided by the embedding procedure in the coronal domain is shown. Similar to panel (a), the flux rope field lines are colored, whereas the surrounding magnetic field is shown as gray curves. Due to the form of the radial advection flow, the trailing part of the flux rope remains in the coronal domain, whereas the front part of the structure is transported approximately self-similarly. 

Using the aforementioned propagation scheme, the magnetic field and plasma properties of the advected embedding region at 0.1 au are used to construct time-dependent inner boundary conditions for EUHFORIA. Prior to inserting the embedding region, a steady-state solar wind condition is established following the standard approach described in \citet{2018Pomoell}. First, based on synoptic magnetogram observations, the PFSS and SCS extrapolation methods \citep{mcgregor} are applied to determine the background coronal magnetic field up to 0.1 au. Subsequently, the empirical formulations detailed in \citet{2018Pomoell} are utilized to generate all associated MHD variables for the background solar wind at the inner radial boundary of EUHFORIA at 0.1 au. Driven by this boundary data, a steady-state (in the co-rotating frame) solar wind solution  is achieved by evolving the MHD simulation for a standard duration of 14 days.

\section{Model validation}\label{validation}
We validate the FROST modeling framework using the well-studied CME event of 2013 April 11. The CME erupted from the Sun on 2013 April 11 at around 06:50 UT and reached Earth on 2013 April 13 at 12:54 UT \citep{Vemareddy_2015}. The magnetic and geometric properties of the CME has been extensively studied in \citet{Sarkar_2020} and the Sun-to-Earth evolution of the CME for this event has been previously simulated using a spheromak CME model in EUHFORIA \citep{Sarkar_2024}. 

We use the CME properties observationally determined by \citet{Sarkar_2020} to constrain the magnetic and geometric properties of the toroidal flux rope to be used in this work. Details on constraining the different properties of the toroidal flux rope using observational proxies are described below.

\subsection{Constraining the 
flux rope geometric properties}

\begin{figure*}[t!]
\begin{center}
\includegraphics[width=\textwidth,clip=]{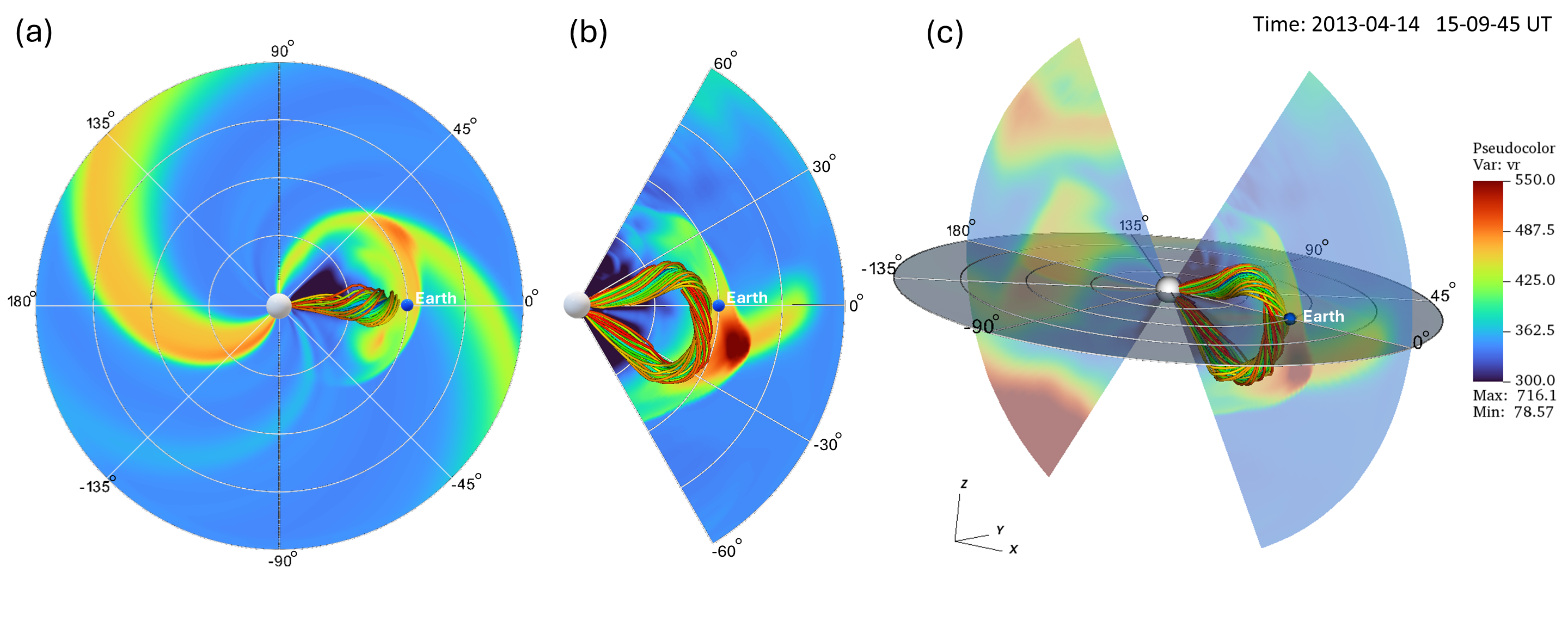}
\caption{Visualization of the flux rope magnetic structure in the heliospheric model Run 1 approximately at the time of arrival of the FR at Earth. Panel (a) shows the view from above the equatorial plane, panel (b) shows the meridional plane containing Earth, and panel (c) shows an arbitrary viewing angle. The white sphere is plotted with a radius equal to the inner boundary of the simulation at 0.1 au, while the full extent of the domain reaches out to 2 au. The dynamic heliosphere is depicted by the background colors, which indicate the solar wind speed. The bundle of twisted lines illustrates the three-dimensional magnetic flux rope structure propagating away from the Sun.
}\label{3d_visulalisation}
\end{center}
\end{figure*}

The CME event on 2013 April 13 was well observed by the SOHO and twin STEREO spacecraft. Employing a graduated cylindrical shell (GCS) \citep{gcs1} model to the multi-vantage point white-light observations from LASCO/C2 onboard SOHO and STEREO-A \& B/COR2 coronagraphs, \citet{Sarkar_2020} reported the near-Sun tilt angle of the CME to be $70\pm 10^{\circ}$. Further, comparing the in-situ observations at L1 with ensemble modeling results obtained using INFROS, \citet{Sarkar_2020} identified the best-fit propagation direction as $0^{\circ}$ longitude and $-10^{\circ}$ latitude within the uncertainty limit of GCS. We use these data to constrain the input tilt and insertion direction of the toroidal flux rope. Notably, the GCS reconstruction for this event yields an uncertainty range of 
$\pm 10^\circ$ for the propagation direction in both longitude and latitude, which we have accounted for by studying the 
model outputs at multiple virtual spacecraft distributed across the uncertainty 
sector (see Section~\ref{results}).  

We specify the major ($R_\mathrm{FR}$) and minor ($a_\mathrm{FR}$) radius of the torus using the parameters obtained from the GCS fitting (see the Appendix for more details). Utilizing multi–viewpoint observations, \citet{Sarkar_2020} fitted the GCS model to the CME flux rope, identified as a bright, curved loop-like structure most clearly seen in STEREO-B base-difference images (see Figure~\ref{shock_fitting}b). Averaging the fitting results obtained at different observation times of the CME within the COR2 field of view yields values of the GCS flux rope aspect ratio, $\kappa$, and half-angle, $\alpha$, of $0.22\pm0.02$ and $28\pm5^{\circ}$, respectively. As described in the Appendix, these average parameter values provide an estimate for the torus geometry, resulting in the major radius $R_\mathrm{FR} = 0.47d$ and minor radius $a_\mathrm{FR} = 0.19d$, with $d$ the heliocentric distance to the torus center. This results in the leading-edge apex height of the torus to be $h_\mathrm{torus} = d + R_\mathrm{FR} + a_\mathrm{FR} = 1.66 d$ for a given distance $d$. Notably, the major and minor radii of the torus being linear functions of $d$, as mentioned above, arise from the assumption that the flux rope evolves self-similarly. 


We further constrain the radius of the envelope containing the draping field around the torus. The leading edge of the envelope, where the draping field transitions to the purely radial external field, can be approximated to be at a  height close to the shock front ahead of the CME. However, it is noteworthy that the transition of the magnetic field at a shock is  abrupt rather than smooth as given by the draping. While the true transition to a purely radial external field may extend slightly beyond the shock height, its precise location is difficult to resolve from white-light observations alone. Therefore, as a physically grounded upper limit, the observed shock height can be used as a proxy to set the leading edge height of the envelope. We therefore perform ellipsoidal fitting to the white-light shock structure of the event to estimate the height of the CME-driven shock, as shown in Figure \ref{shock_fitting}. The geometrical fittings performed on the white-light structures within the COR2 field of view yield an average ratio of the shock height to the GCS leading-edge height of 1.19. Assuming self-similar expansion, this yields for the shock height $h_{\text{shock}} = 1.19 h_{\text{torus}} = 1.98 d$ at a given distance $d$. Using this relation, we determine the heliocentric distance of the center of the toroidal embedding $d_E$ to be $10.85 R_\odot$ for the initial state of the coronal magnetic field by equating the shock height with the inner boundary height (0.1 au) of the heliospheric model. Thus, the embedding of the flux rope into the ambient wind is performed such that the apex height of the envelope reaches $r = 0.1$ au. This also provides the values of $R_\mathrm{FR}$ and $a_\mathrm{FR}$ at $d_E = 10.85 \, R_\odot$ as $5.10 R_\odot$ and $2.06 R_\odot$, respectively. 


We estimate the shock arrival time at 0.1 au by extrapolating the observed height–time profile of the shock obtained from ellipsoidal fitting. Based on this arrival time, we determine the timestamp corresponding to the initial state of the coronal magnetic field. We refer to this timestamp as the insertion time of the CME in the heliospheric simulation, which includes the draping field as well as the flux rope. The information on the insertion time, as well as the major and minor radii of the torus and the radius of the envelope region, estimated based on the value of $d_E \approx 11 R_\odot$, are listed in Table \ref{table1}, with the values rounded to the nearest integer.
\begin{deluxetable}{ll}[!h]
\tablecaption{CME input parameters for simulation Run 1 \label{table1}}
\tablehead{
\colhead{Parameter} & \colhead{Value} 
}
\startdata
Insertion time & 2013-04-11T11:12:00 \\
Speed ($v_\mathrm{radial}$) & 650\;km s$^{-1}$ \\
Longitude $\phi_E$ & 0$^{\circ}$ \\
Latitude $\theta_E$ & -10$^{\circ}$ \\
Toroidal embedding distance $d_E$  & 11\;\RSun \\
Flux rope major radius $R_\mathrm{FR}$ & 5\;\RSun \\
Flux rope minor radius $a_\mathrm{FR}$& 2\;\RSun \\
Envelope minor radius $a_\mathrm{E}$ & 3.5\;\RSun \\
Helicity sign & -1 (Left handed) \\
Tilt $\tau_E$ &  70$^{\circ}$ \\
Toroidal flux & $2\times10^{13}$ Wb \\
Density & $1\times10^{-18}$\;kg m$^{-3}$\\
Temperature & $0.8\times10^{6}$\;K \\
\enddata
\end{deluxetable}
\begin{deluxetable}{lll}[!b]
\tablecaption{Model runs with different density and speed values for the CME \label{table2}}
\tablehead{
\colhead{Run number} & \colhead{Density} & \colhead{Speed ($v_\mathrm{radial}$)}  
}

\startdata
Run 1      & 1$\times10^{-18}$\;kg m$^{-3}$ & 650\;km s$^{-1}$\\
Run 2      & 1$\times10^{-18}$\;kg m$^{-3}$ & 750\;km s$^{-1}$\\
Run 3      & 5$\times10^{-18}$\;kg m$^{-3}$ & 650\;km s$^{-1}$\\
Run 4      & 5$\times10^{-18}$\;kg m$^{-3}$ & 750\;km s$^{-1}$\\
Run 5      & 10$\times10^{-18}$\;kg m$^{-3}$ & 650\;km s$^{-1}$\\
Run 6      & 10$\times10^{-18}$\;kg m$^{-3}$ & 750\;km s$^{-1}$\\
\enddata
\end{deluxetable}

\begin{figure}[t!]
	\begin{center}
		\includegraphics[width=0.48\textwidth,clip=]{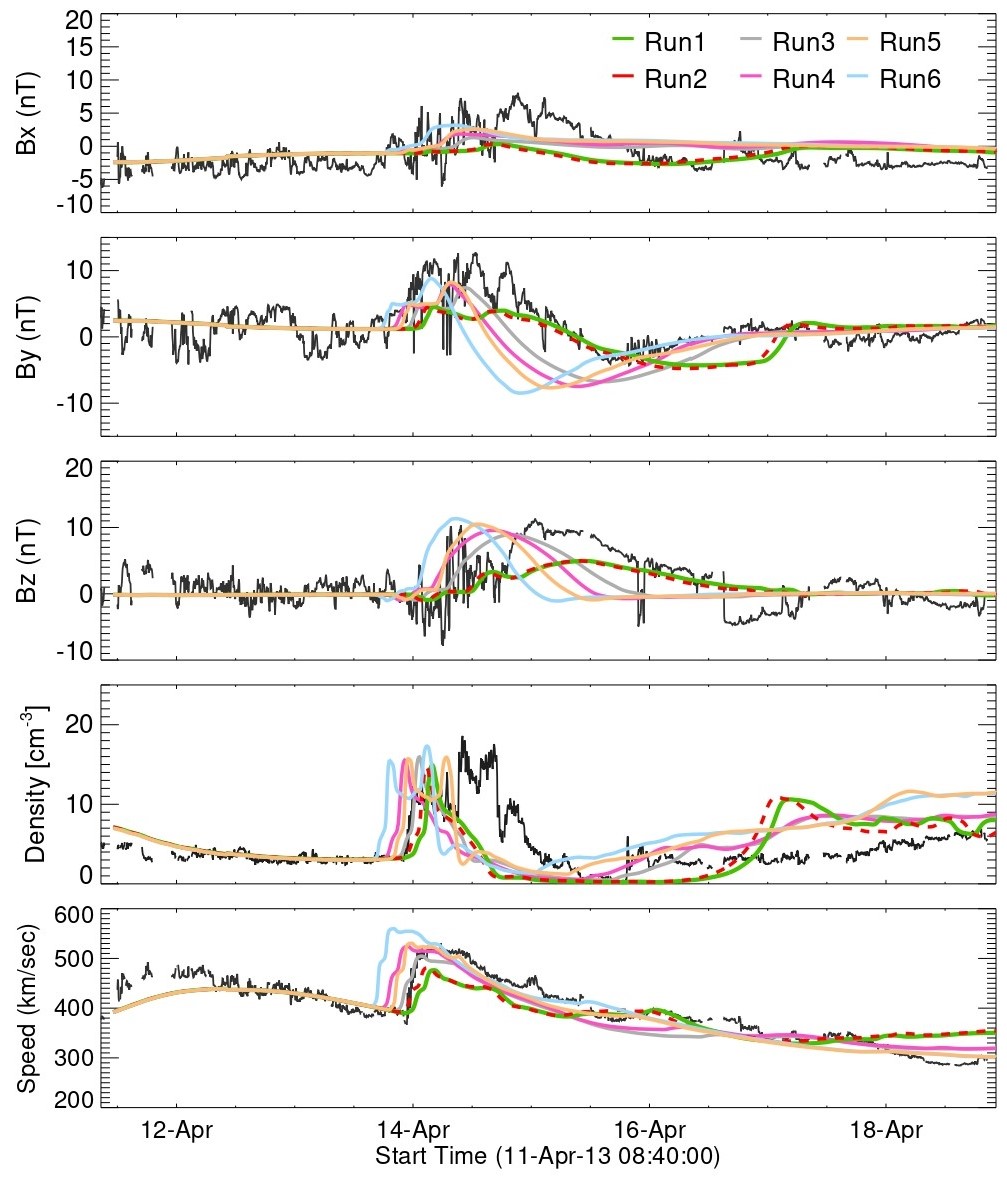}
		\caption{Comparison of the model results for all the six runs with the observed in situ
			plasma properties. The black solid curves denote the observation from OMNI, and the color coded solid curves are the model output at the location of Earth.}\label{all_run_comparison}
	\end{center}
\end{figure}

\begin{figure*}[t!]
	\begin{center}
		\includegraphics[width=\textwidth,clip=]{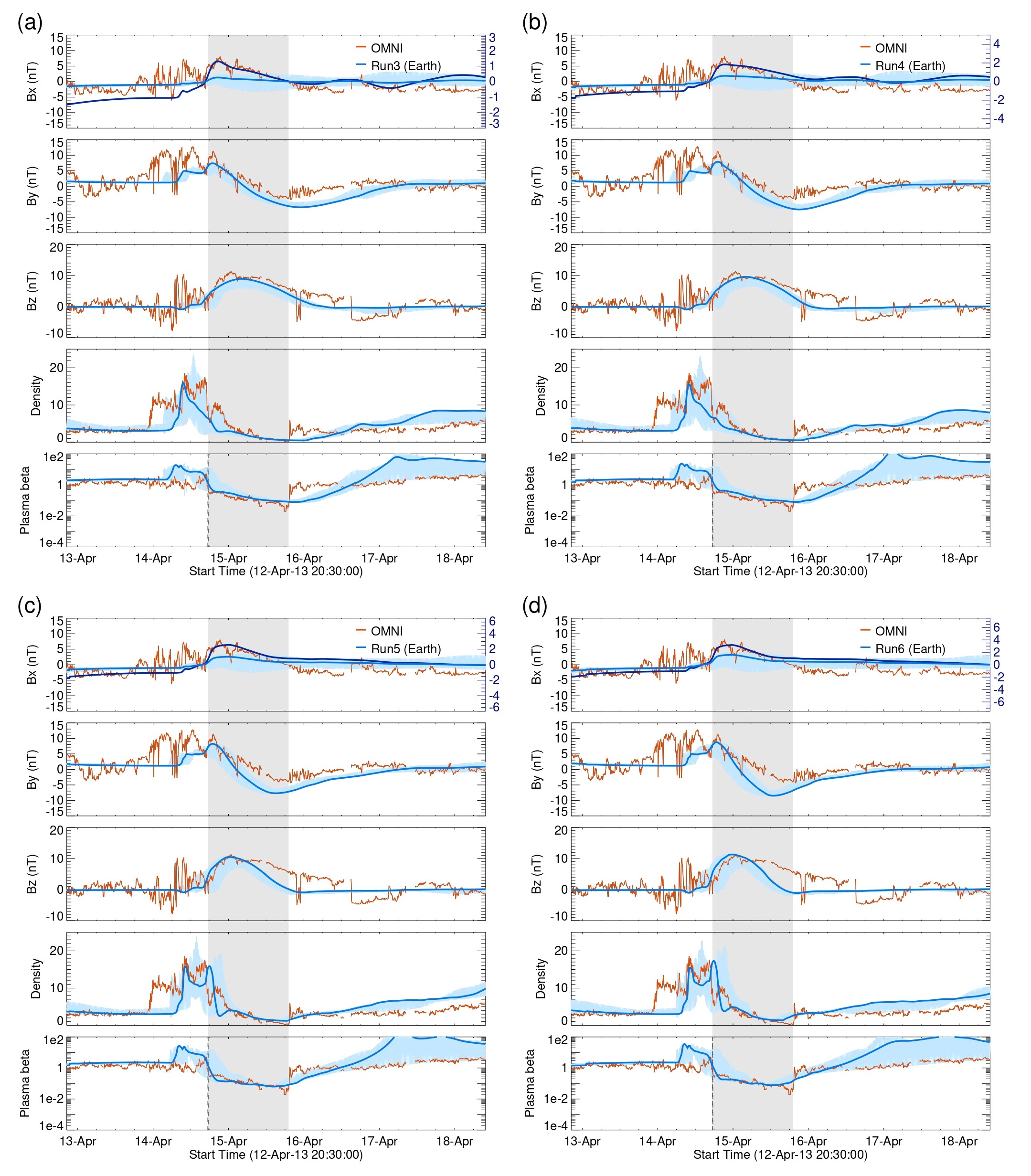}
		\caption{Comparison of the observed magnetic field vectors, density and plasma beta of the ICME and the time-shifted model results for Run 3 (panel a), Run 4 (panel b), Run 5 (panel c) and  Run 6 (panel d), respectively. }\label{time_shifted_profiles}
	\end{center}
\end{figure*}
\subsection{Constraining the flux rope magnetic properties}
The orientation of the axial magnetic field of the flux rope near the Sun was determined to be northward directed by \citet{Sarkar_2020}, based on observations of the magnetic polarities underlying the two footpoints of the sigmoidal structure associated with the eruption. Furthermore, using the GCS reconstruction, \citet{Sarkar_2020} reported the inclination angle of the flux rope to be $70^\circ$ (measured anticlockwise with respect to the solar equatorial plane) at $\approx 10\,R_\odot$. We use this estimate as the tilt angle $\tau_E$ of the torus when performing the embedding, i.e., for which $d_E \approx 11\,R_\odot$, while constraining its axial direction such that the magnetic connectivity of the two anchoring legs of the flux rope is oriented from south to north when the flux rope emerges from the heliospheric inner boundary.

To estimate the toroidal flux of the torus, we adopt a method similar to that used in our previous work to observationally constrain a spheromak \citep{Sarkar_2024}. As part of this method, we equate the magnetic field strength at the axis of the torus to the axial field strength ($B_0$) of the CME at distance $d_E$, obtained from the Flux Rope Eruption Data (FRED) technique \citep{IAU_Gopalswamy}. Conceptually, FRED uses the information on reconnection flux \citep{Kazachenko_2017,2018Gopal} and a Lundquist
flux rope model \citep{lundquist} constrained with the geometric properties of the flux rope to estimate $B_0$ at a given distance. Applying the FRED technique, \citet{Sarkar_2020} reported the $B_0$ of the associated CME at 10 $R_\odot$ as 5200 nT, which corresponds to an approximate $B_0$ of 4400 nT at $d_E = 10.85~R_\odot$, assuming self-similar expansion. Using the estimated $B_0$ value, we determine the toroidal flux to be $2 \times$10$^{13}$ Wb. Further, we assign the chirality of the toroidal flux rope as left-handed, based on the observational study on this event by \citet{Sarkar_2020}. The magnetic properties of the flux rope, are summarized in Table \ref{table1}.

\subsection{Constraining the speed and density}
Based on our analysis using the GCS reconstruction method, we estimate the average near-Sun speed ($v_\mathrm{3D}$) of the CME to be in the range of approximately 800 km s$^{-1}$ to 900 km s$^{-1}$. Notably, the speed ($v_\mathrm{3D}$) obtained from the GCS contains contribution from translational speed ($v_\mathrm{radial}$) and the expansion speed of the structure. Since EUHFORIA models the CME expansion in a self-consistent manner, only the translational component of the velocity is required as an input parameter.

Following the formulation introduced by \citet{2019scolini}, the translational component $v_\mathrm{radial}$ can be expressed as
\begin{equation}
v_{\mathrm{radial}}=\frac{{v_\mathrm{3D}}}{1+\kappa},
\end{equation}
where $\kappa$ represents the aspect ratio obtained from the GCS fitting procedure. Using an aspect ratio of $\kappa = 0.22$ \citep{Sarkar_2020}, the corresponding value of $v_{\mathrm{radial}}$ is estimated to be in the range of 650 km s$^{-1}$ to 750 km s$^{-1}$.

We further use density estimates previously considered by \citet{Sarkar_2024}, which are in the order of 1$\times10^{-17}$\;kg m$^{-3}$ to 1$\times10^{-18}$\;kg m$^{-3}$. To account for the uncertainties in the speed and density specification, we perform multiple simulations runs as varying these parameters and listed in Table \ref{table2}.

The final required input parameter of the CME, the plasma temperature, is set to $0.8 \cdot 10^{6}$\;K, which is commonly used in studies employing EUHFORIA, and is for simplicity assumed to be uniform inside the CME \citep{2018Pomoell}. To construct the ambient heliospheric solar wind, we utilize the  background solar wind solution for this event as provided by \citet{Sarkar_2024}. 
In the following section, we discuss the results of the simulations based on the different runs listed in Table \ref{table2}. 

\begin{deluxetable*}{lllllll}[!t]
\tablecaption{Quantitative assessments of the model results as obtained from Runs 3 to 6. The Pearson correlation coefficients between the observed and modeled magnetic vectors $B_x$, $B_y$, and $B_z$ within the flux rope boundary, are denoted as $CC_{B_x}$, $CC_{B_y}$, and $CC_{B_z}$, respectively. $\Delta{B_x}$, $\Delta{B_y}$ and $\Delta{B_z}$ refer to the normalized RMS error in the modeled $B_x$, $B_y$, and $B_z$ respectively. \label{table3}}
\tablehead{
\colhead{Run number} & \colhead{    $CC_{B_x}$} & \colhead{    $CC_{B_y}$} & \colhead{    $CC_{B_z}$} & \colhead{    $\Delta_{B_x}$ (nT)} & \colhead{    $\Delta_{B_y}$ (nT)} & \colhead{    $\Delta_{B_z}$ (nT)}
}

\startdata
Run 3      &  0.93 & 0.95 & 0.86 & 0.42  & 0.20 & 0.14 \\
Run 4      &  0.91 & 0.95 & 0.85 & 0.35  & 0.23 & 0.12 \\
Run 5      &  0.64 & 0.92 & 0.87 & 0.31 & 0.31 & 0.20 \\
Run 6      &  0.72 & 0.92 & 0.82 & 0.27  & 0.37 & 0.27 \\
\enddata
\end{deluxetable*}

\subsection{Model results}\label{results}
Figure \ref{3d_visulalisation} shows a 3D visualization of the CME flux rope after approximately 76 hours of MHD evolution in the heliospheric domain. Notably, the flux rope structure remains coherent throughout its evolution, likely due to the isolated propagation of the CME in the absence of any interaction with high-speed streams or other CMEs. Due to the novel insertion technique of the flux rope, its two legs remain attached to the inner boundary of the heliospheric domain, mimicking a canonical CME morphology consistent with the observations of bi-directional electron flows \citep[e.g.,][]{Gosling1990}. Importantly, the modeled flux rope approximately maintains its axial tilt throughout the propagation, in contrast to the significant rotation (up to $90 ^{\circ}$) experienced by the spheromak model studied for the same event under an identical background solar wind in the previous work by \citep{Sarkar_2024}.

We compare the model outputs at the position of Earth extracted from each simulation run listed in Table \ref{table2} with the observed in situ profile of the ICME recorded by OMNI (see Figure \ref{all_run_comparison}). Notably, the profiles of all magnetic field components ($B_x$, $B_y$, and $B_z$) within the ICME show very good agreement with the observed time-series. The magnitude of the components is also well captured particularly in Runs $3-6$, whereas the low-density simulation runs (Runs 1 and 2) underestimate the field strength. The lower magnetic field strength at the spacecraft location observed in the simulation runs employing a lower mass density in the CME is consistent with the findings of \citet{Sarkar_2024}. The figure also shows that different simulation runs yield varying ICME arrival times at 1 au, deviating from the observed arrival time, with absolute errors in the associated shock arrival time ranging from approximately 30 minutes to 4 hours.

Therefore, to compare the predicted magnetic field vectors of the CME with the observed ones, we apply a constant time shift to each model run, aligning the arrival of the modeled and observed flux ropes. This alignment is achieved by matching the plasma beta profiles at the point where the plasma beta drops below unity in both the observed and modeled time series, as indicated by the dashed vertical black lines in Figure~\ref{time_shifted_profiles}, which mark the starting boundary of the flux rope. Since the magnetic field strength of the ICME is underestimated in Runs 1 and 2, we show only Runs $3-6$ in Figure \ref{time_shifted_profiles}. The applied time shifts for Runs 3, 4, 5, and 6 are 8.2, 11.5, 11.0, and 16.5 hours, respectively. The solid blue lines in Figure \ref{time_shifted_profiles} represent the time-shifted model results at Earth and the shaded light blue regions denote the same at various virtual spacecraft (SC) located within $\pm 10^{\circ}$ in both longitudinal and latitudinal directions with respect to the Earth. The close match between the modeled and observed magnetic field components of the flux rope, indicated by the gray-shaded region, is now even more evident. Particularly, the match is excellent for $B_y$, and $B_z$.
Although the magnitude of $B_x$ is underestimated for all the runs, comparing the $B_x$ profiles using adjusted scales (represented by the solid dark blue lines in the top panels of Figure \ref{time_shifted_profiles}a–d) highlights a strong qualitative agreement between the observed and modeled data.

However, in the model results, the virtual spacecraft at 1 au detects a high plasma beta after crossing the flux rope (from approximately 16 April 2013, 12:00 UT onward) that is not present in the observations. In the model, the high plasma beta results from the presence of a current sheet in the model that is formed between the two legs of the flux rope.

In order to assess the model results quantitatively, we evaluate the Pearson correlation coefficient (CC) between the observed and modeled magnetic vectors and estimate the normalized RMS error ($\Delta {B_{x,y,z}}$) in the modeled $B_x$, $B_y$, and $B_z$ components, where $\Delta_{B_{x,y,z}}$ is defined as
\begin{equation}
\Delta{B_{x,y,z}}= \frac{1}{\left| B_{x,y,z} \right|^\mathrm{obs}_\mathrm{max}} \sqrt{ \frac{\sum^{N}_{i=1}({B_{x,y,z}^\mathrm{model}}(t_i)-{B_{x,y,z}^\mathrm{obs}}(t_i))^2}{N}}
\end{equation}
%
%
Here, ${B_{x,y,z}^\mathrm{obs}}(t_i)$ and ${B_{x,y,z}^\mathrm{model}}(t_i)$ represent the observed and modeled magnetic fields within the ICME boundary as shown by the gray shaded region in Figure \ref{time_shifted_profiles}, respectively, at time $t_i$, sampled over $N$ data points of the modeled magnetic field. ${\left| B_{x,y,z} \right|^\mathrm{obs}_\mathrm{max}}$ denotes the maximum magnitude of the observed components. To match the 10-minute temporal cadence of EUHFORIA model outputs, the 1-minute in-situ observations were processed using a 10-minute running average and then downsampled to 10-minute intervals within the duration of the observed ICME boundary ($\approx$ 1 day). The number of sampled data points ($N \approx 150$) ensures the statistical 
significance of the calculated CC, yielding a $p$-value of zero (or $p < 0.001$) across all runs. The results are listed in Table \ref{table3}, which quantitatively confirm the excellent match between the observed and modelled magnetic field components for Runs $3-6$. Notably, the CC for all components of the magnetic field are greater than or equal to 0.85 for Run3 and Run4. The CC is high for Run5 and Run6 as well, except for the $B_x$ component, which shows a comparatively lower value than Run3 and Run4. This is due to the mismatch of $B_x$ towards the back-half of 
the flux-rope (see the solid dark blue lines in the top panels of Figure~\ref{time_shifted_profiles}(c) \& (d)). However, the magnitude of $B_x$ is comparatively well captured in Run5 and Run6, resulting in a lower 
$\Delta B_x$ compared to Run3--4. A mismatch towards the back-half of the flux-rope can also be seen for the $B_y$ and $B_z$ profiles for Run5 and 
Run6, which results in higher $\Delta B_y$ and $\Delta B_z$ for these runs as compared to Run3--4.

\section{Discussion and conclusions}\label{conclusion}

In this work, we implemented a new flux rope model, FROST, for use with the global heliospheric MHD simulation EUHFORIA, introducing two key methodological advances beyond the current state-of-the-art space weather forecasting tools. First, the inclusion of the embedding magnetic field ensures continuity of the normal component of the magnetic field at the flux rope boundary while preserving the original flux-rope solution. Second, the flux-rope insertion technique---using a self-similar advection speed while anchoring the two legs of the flux rope below the inner boundary of the heliospheric simulation, produces a canonical CME configuration consistent with observations of bi-directional electron flows \citep[e.g.,][]{Gosling1990}, which are not reproduced by fully inserted spheromak or toroidal flux-rope models.


Our study validated the FROST modeling framework for the CME event of 2013 April 11 and showed an excellent agreement between the observed and modeled magnetic field profiles of the ICME at 1 AU. Importantly, the higher-density runs, performed using two ICME speeds (650 and 750 km s$^{-1}$), successfully reproduce both the magnitude and orientation of all the magnetic field components of the ICME flux rope. In particular, the magnitude of the modeled magnetic field components did not vary significantly with the CME speed across all performed runs. In contrast, we found that mass density appears to be a critical parameter to obtain realistic magnetic field magnitudes at 1 au. Therefore, rather than using ad hoc values, the flux-rope input density must be accurately determined using density estimation techniques \citep{2021Mannuela}, which use CME mass and volume measurements at specific coronal heights and scale the density to 0.1 au assuming self-similar expansion. Furthermore, this result implies that the assumption of employing a uniform density inside the flux-rope and embedding may not be optimal and requires additional future work to quantify in detail. 

It is noteworthy that the arrival time errors for the flux-rope leading edge (up to 16.5 hours) in this study are significantly larger than those obtained for the shock arrival (up to 4 hours). This discrepancy stems from the modeled sheath duration being shorter than the observed one at 1 au. A possible source of this discrepancy may be due to the parameter values chosen in constructing the embedding region that may not accurately capture the size of the sheath at the point of injection. This underestimation is possibly exacerbated by setting the inner boundary of the heliospheric simulation domain at 0.1 au, which inherently bypasses the region of self-consistent sheath formation below this distance. While other physical factors inherent to sheath formation and expansion may also play a role, addressing them remains beyond the scope of the current paper. In future work, we plan to include coronal simulations of sheath formation by extending the simulation domain to include a larger portion of the coronal domain.  

The spheromak model used for the same event under identical solar wind conditions was found to rotate by up to $\approx 90^\circ$ during its propagation toward Earth, resulting in inconsistent in situ magnetic profiles and large uncertainties in predicting $B_z$ \citep{Sarkar_2024}. In contrast, the flux-rope model implemented in this work does not exhibit such significant rotation during its heliospheric propagation. This stability makes it a more reliable tool for predicting $B_z$. Moreover, constructing the flux rope solution in FROST requires only a few minutes of wall-clock time on a single CPU core and therefore does not appreciably add to the computational cost of running the heliospheric model. These highlight its strong potential for advancing space-weather forecasting. Nevertheless, a larger set of events in different background solar wind conditions must be examined in future work to fully assess its robustness and general applicability.

The embedded magnetic field around the flux rope not only prevents numerical artifacts at the flux rope boundary but also provides an effective means of modeling the sheath magnetic field formed due to field line draping. While this article focuses on validating the flux rope properties, we plan to explore the formation and evolution of sheath regions using our new approach in forthcoming studies. 

\section*{Acknowledgments}
RS acknowledges support from the Research Council of Finland Grant 350015. JP acknowledges Research Council of Finland Projects 343581, 364852 and 370793. EA acknowledges support from the Research Council of Finland (Research Fellow grant number 355659). EK, EA and JP acknowledge the ERC under the European Union's Horizon 2020 Research and Innovation Programme Project 724391 (SolMAG). The work at the University of Helsinki was performed under the umbrella of Finnish Centre of Excellence in Research of Sustainable Space (Research Council of Finland Grant no. 312390, 336807). EK and JP also acknowledge the SOLER project funded by the European Union's Horizon Europe research and innovation programme under grant agreement No 101134999. Open access is funded by Helsinki University Library.

\section*{Appendix}\label{appendix}
\begin{figure}[!h]
\begin{center}
\includegraphics[width=0.41\textwidth]{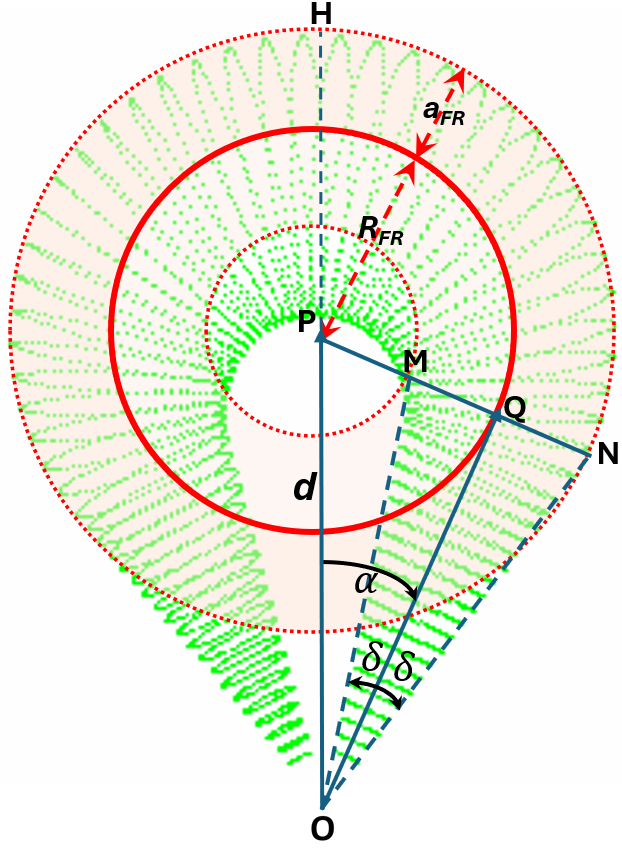}
\caption{Representation of the toroidal geometry (enclosed by the two red dotted circles) constrained with the GCS model (depicted by the green dots) viewed face-on. }\label{GCS_torus}
\end{center}
\end{figure}

The major radius $R_\mathrm{FR}$ and minor radius $a_\mathrm{FR}$ of the toroidal flux rope (see Section \ref{subsec:embedding}) are constrained using the GCS parameters that include the aspect ratio $\kappa$ and half-angle $\alpha$. The half-angle of the GCS croissant is defined as the angle between the axis (\( \overrightarrow{OQ} \)) of one of its conical legs and the central axis ($\overrightarrow{OP} $) along the direction of propagation as shown in Figure \ref{GCS_torus}. The angular half-width of each conical leg is given by $\delta = \angle QON =\angle QOM$ and is related to the aspect ratio via 
\begin{eqnarray}
\kappa=\sin \delta \label{kappa}  
\end{eqnarray}
as defined by \citet{gcs1}. For a given distance $d = |\overrightarrow{OP}|$ , which represents the distance between the origin of the system (typically the center of the Sun) $O$ and the center $P$ of the torus, the line that passes through $P$ and that is perpendicular to \(\overline{OQ}\), is denoted as \(\overline{PN}\). The axis of the conical leg intersects \(\overline{PN}\) at \( Q \). We set \( PQ\)
to be the major radius of the torus,  and depicted as the solid red circle in Figure \ref{GCS_torus}. \(\overline{PN}\) is further intersected by the two edge boundaries (denoted as blue dashed lines) of the conical leg at \( M \) and \( N \). Therefore, we set $QM$ and $QN$ to be the minor radius of the torus, and is constrained by the half-angular extent of the conical legs of the croissant. From these, we obtain 
\begin{eqnarray}
R_{FR}= PQ = d \sin  \alpha
\label{semi_major}
\end{eqnarray}
and
\begin{eqnarray}
a_{FR}= QM = OQ \sin \delta= d \cos \alpha \sin \delta . \label{semi_minor}
\end{eqnarray}
Further using equations \ref{kappa} and \ref{semi_minor}, the minor radius can be expressed as
\begin{eqnarray}
a_{FR}= d \, \kappa \cos \alpha.
\end{eqnarray}
The distance $d$ is constrained such that 
\begin{eqnarray}
OH = d + R_\mathrm{FR} + a_\mathrm{FR} =
d (1 + \sin \alpha + \kappa \cos \alpha)
\end{eqnarray}
where $OH$ is the distance from the origin to the GCS apex.
It is important to note that the point $P$ in this construct is not equal to the point $B$ defined as the center of the GCS generating circle in \citet{gcs1}. Therefore, the constructed torus volume deviates from the volume of the GCS model, as is also seen in the illustration in Figure \ref{GCS_torus}.

\bibliography{references}
\end{document}